\documentclass{llncs}

\usepackage[T2A,T1]{fontenc}
\usepackage[cp1251]{inputenc}
\usepackage{float}
\usepackage{amsmath}
\usepackage{setspace}
\makeatletter

\numberwithin{equation}{section}
\numberwithin{figure}{section}

\usepackage{fancyhdr}
\usepackage{amsfonts,amsmath,amssymb}
\usepackage{algorithm,algpseudocode}
\usepackage{graphicx,caption}
\usepackage{cmap}

\numberwithin{equation}{section}

\makeatletter
\newcommand{\newalgname}[1]{%
  \renewcommand{\ALG@name}{#1}%
}
\newalgname{Algorithm}
\makeatother

\renewcommand{\thefigure}{\thesection.\arabic{figure}}

\begin{document}

\title{ALGORITHMIC COMPUTATION OF~POLYNOMIAL AMOEBAS}

\author{D.~V.~Bogdanov\inst{1} \and A.~A.~Kytmanov\inst{2} \and T.~M.~Sadykov\inst{3}}

\institute{Plekhanov Russian University \\
125993 Moscow, Russia \\ \email{bogdv@rambler.ru} \and Siberian
Federal University \\ Svobodny~79, Krasnoyarsk, 660041, Russia \\
\email{aakytm@gmail.com} \and
Plekhanov Russian University \\
125993 Moscow, Russia \\ \email{Sadykov.TM@rea.ru} }

\maketitle

\begin{abstract}
We present algorithms for computation and visualization
of~amoebas, their contours, compactified amoebas and sections of
three-dimensional amoebas by two-dimensional planes. We also
provide a~method and an algorithm for the computation
of~polynomials whose amoebas exhibit the most complicated topology
among all polynomials with a~fixed Newton polytope. The presented
algorithms are implemented in computer algebra systems Matlab~8
and Mathematica~9.


\end{abstract}

\section{Introduction}

The amoeba of~a~multivariate polynomial in several complex
variables is the Reinhardt diagram of~its zero locus in the
logarithmic scale with respect to each of~the
variables~\cite{Viro}. The term <<amoeba>> has been coined
in~\cite[Chapter~6]{GKZ} where two competing definition of~the
amoeba of~a~polynomial have been given: the affine and the
compactified versions. Both definitions are only interesting in
dimension two and higher since the amoeba of~a~univariate
polynomial is a~finite set that can be explored by a~variety
of~classical methods of~localization of~polynomial roots.

The geometry of~the amoeba of~a~polynomial carries much
information on the zeros of~this polynomial and is closely related
to the combinatorial structure of~its Newton polytope (see
Theorem~\ref{3thmfptestimate}). Despite loosing half of~the real
dimensions, the image of~the zero locus of~a~polynomial in the
amoeba space reflects the relative size of~some of~its
coefficients.

From the computational point of~view, the problem of~giving
a~complete geometric or combinatorial description of~the amoeba
of~a polynomial is a~task of~formidable complexity~\cite{Purbhoo},
despite the substantial recent progress in this
direction~\cite{Theobald,TW,Rullgaard}. The number of~connected
components of~an amoeba complement as a~function of~the
coefficients of~the polynomial under study is still to be explored
by means of~the modern methods of~computer algebra. In particular,
the conjecture by M.Passare on the solidness of~maximally sparse
polynomials (see Definition~\ref{def:solidAmoeba}) remains open
for a~long time.

Amoebas can be computed and depicted by means of~a~variety
of~approaches and methods
(see~\cite{Forsberg,Purbhoo,Nilsson,Theobald,Johansson,TW} and the
references therein). The present paper is meant to move forward
the art of computing and drawing complex amoebas of algebraic
hypersurfaces. A~special attention paid to the geometrically most
interesting and computationally most challenging case of optimal
hypersurfaces (see Definition~\ref{def:optimalAmoeba}). We expose
methods and algorithms for computation and visualization
of~amoebas of~bivariate polynomials, their contours and
compactified versions. The developed algorithms are used in higher
dimensions for depicting sections of~amoebas of~polynomials in
three variables. The main focus of~the paper is on polynomials
whose amoebas have the most complicated topological structure
among all polynomials with a~given Newton polytope (see
Definition~\ref{def:optimalAmoeba}). We provide an algorithm for
explicit construction of~such polynomials.

The presented algorithms are implemented in the computer algebra
systems Matlab~8~(64-bit) and Wolfram Mathematica~9~(64-bit). All
examples in the paper have been computed on Intel Core~i5-4440 CPU
clocked at~3.10~GHz with~16~Gb RAM under MS~Windows~7
Ultimate~SP1.

\section{CONVEX POLYTOPES, CONES AND AMOEBAS: DEFINITIONS AND
PRELIMINARIES}

Let $p\left(x\right)$ be a~polynomial in $n$ complex variables:
\[
p\left(x_{1},\ldots,x_{n}\right)=\sum_{\alpha\in
A}c_{\alpha}x^{\alpha} = \sum_{\alpha\in
A}c_{\alpha_{1}\ldots\alpha_{n}}x_{1}^{\alpha_{1}}\cdot\ldots\cdot
x_{n}^{\alpha_{n}},
\]
where $A\subset\mathbb{Z}^{n}$ is a~finite set.

\begin{definition}\rm\textit{The Newton polytope}~$\mathcal{N}_{p\left(x\right)}$
of a~polynomial~$p\left(x\right)$ is the convex hull of~the
set~$A$ of~its exponent vectors.
\end{definition}

\begin{definition}\rm\textit{The recession cone} of~a~convex set~$M$
is the set-theoretical maximal element in the family of~convex
cones whose shifts are contained in~$M$.
\end{definition}

\begin{definition}\rm\label{def:Amoeba}\textit{The amoeba}~$\mathcal{A}_{p\left(x\right)}$
of a~polynomial~$p\left(x\right)$ is the image of~its zero locus
under the map
\[
\mathrm{Log}:\;\left(x_{1},\ldots,x_{n}\right)\longmapsto
\left(\ln\left|x_{1}\right|,\ldots,\ln\left|x_{n}\right|\right).
\]
\end{definition}

The connected components of~the complement to the
amoeba~$\mathcal{A}_{p\left(x\right)}$ are convex and in bijective
correspondence with the expansions of~the rational function
$1/p(x)$ into Laurent series centered at the origin~\cite{FPT}.
The next statement shows how the Newton
polytope~$\mathcal{N}_{p\left(x\right)}$ is reflected in the
geometry of~the
amoeba~$\mathcal{A}_{p\left(x\right)}$~\cite[Theorem~2.8 and
Proposition~2.6]{FPT}.

\begin{theorem}\rm (See~\cite{FPT}.) \label{3thmfptestimate}
Let~$p\left(x\right)$ be a~Laurent polynomial and let $\left\{
M\right\}$ denote the family of~connected components of~the amoeba
complement~$^{c}\!\mathcal{A}_{p\left(x\right)}$. There exists an
injective function
$\nu:\{M\}\rightarrow\mathbb{Z}^{n}\cap\mathcal{N}_{p\left(x\right)}$,
such that the cone that is dual to the
polytope~$\mathcal{N}_{p\left(x\right)}$ at the
point~$\nu\left(M\right)$ coincides with the recession cone
of~$M$. In particular, the number of~connected components
of~$^{c}\!\mathcal{A}_{p\left(x\right)}$ cannot be smaller than
the number of~vertices of~the
polytope~$\mathcal{N}_{p\left(x\right)}$ and cannot exceed the
number of~integer points
in~$\mathcal{N}_{p\left(x\right)}$.\end{theorem}

Thus the amoeba of~polynomial $p(x)$ in $n\geqslant2$ variables is
a closed connected unbounded subset of~$\mathbb{R}^{n}$ whose
complement consists of~a~finite number of~convex connected
components. Besides, a~two-dimensional amoeba has <<tentacles>>
that go off to infinity in the directions that are orthogonal to
the sides of~the polygon~$\mathcal{N}_{p\left(x\right)}$ (see
Fig.~\ref{img:amoeba12}, \ref{img:amoeba34},
\ref{img:big_amoeba}).

The two extreme values for the number of~connected components
of~an amoeba complement are of~particular interest.

\begin{definition}\rm (See~\cite[Definition~2.9]{FPT}.) \label{def:optimalAmoeba}
An algebraic hypersurface
$\mathcal{H}\subset(\mathbb{C}^{*})^{n}$, $n\!\geqslant\!2$, is
called \textit{optimal} if the number of~connected components
of~the complement of~its amoeba~$^{c}\!\mathcal{A}_{\mathcal{H}}$
equals the number of~integer points in the Newton polytope of~the
defining polynomial for~$\mathcal{H}$. We will say that
a~polynomial (as well as its amoeba) is optimal if its zero locus
is an optimal algebraic hypersurface.
\end{definition}

In other words, an algebraic hypersurface is optimal if the
topology of~its amoeba is as complicated as it could possibly be
under the condition that the Newton polytope of~the defining
polynomial is fixed. The other extreme case of~the topologically
simplest possible amoeba is defined as follows.

\begin{definition}\rm (See \cite{PST}.) \label{def:solidAmoeba}
An algebraic hypersurface
$\mathcal{H}\subset(\mathbb{C}^{*})^{n}$, $n\geqslant2$, is called
\textit{solid}, if the number of~connected components of~its
amoeba complement $^{c}\!\mathcal{A}_{\mathcal{H}}$ equals the
number of~vertices of~the Newton polytope of~its defining
polynomial~$\mathcal{H}$.\end{definition}

Thus the solid and the optimal amoebas are the endpoints of~the
spectrum of~amoebas of~polynomials with a~given Newton polytope.
Of course there exist plenty of~optimal (as well as solid) amoebas
defined by polynomials with a~given Newton polytope. In fact, both
sets of~amoebas regarded as subsets in the complex space
of~coefficients of~defining polynomials, have nonempty interior.

In the bivariate case, an amoeba is solid if and only if all
of~the connected components of~its complement are unbounded and no
its tentacles are parallel. A~two-dimensional optimal amoeba has,
on the contrary, the maximal possible number of~bounded connected
components in its complement and the maximal possible number
of~parallel tentacles.

The functional dependency of~the topological type of~the amoeba
$\mathcal{A}_{p\left(x\right)}$ on the coefficient of~its defining
polynomial $p\left(x\right)$ is complex and little understood at
the present moment. A~sufficient condition for the amoeba
of~a~polynomial to be optimal is that it satisfies a~<<natural>>
system of~partial differential equations of~hypergeometric
type~\cite{DickensteinSadykovDoklady,DS} while the support of~the
polynomial in question is complex enough~\cite{ST}. In
Section~\ref{optimal_poly} we expose an algorithm for the
computation of~the hypergeometric polynomial with the prescribed
Newton polytope.

\begin{example}\rm\label{exm:imgAmoeba} Let $\mathcal{N}$ denote
the convex hull of~the set of~lattice points
$\{\left(0,0\right),\left(1,0\right),$
$\left(0,2\right),\left(2,1\right)\}$, see Fig.~\ref{img:Newton}.
This polygon will appear in several examples that follow and has
been chosen as one of~the simplest polygons that contain an inner
integer point as well as an integer point in the relative interior
of its edge.
\begin{figure}[ht!]
\begin{centering}
\includegraphics[width=0.25\columnwidth]{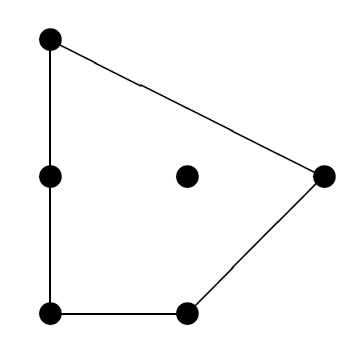}
\par\end{centering}

\caption{The support of~the polynomials in
Example~\ref{exm:imgAmoeba}} \label{img:Newton}
\end{figure}
Fig.~\ref{img:amoeba12} and~\ref{img:amoeba34} show the amoebas
of~the four bivariate polynomials
$p_{1}\left(x,y\right)=1+x+y+xy+y^{2}+x^{2}y$,
$p_{2}\left(x,y\right)=1+x+3y+xy+y^{2}+x^{2}y$,
$p_{3}\left(x,y\right)=1+x+y+4xy+y^{2}+x^{2}y$,
$p_{4}\left(x,y\right)=1+x+3y+4xy+y^{2}+x^{2}y$ whose Newton
polygons coincide with~$\mathcal{N}$. The complement of~the solid
amoeba in Fig.~\ref{img:amoeba12}~(left) consists of~the four
unbounded connected components with two-dimensional recession
cones. The complement to the optimal amoeba in
Fig.~\ref{img:amoeba34}~(right) comprises six connected
components: the four unbounded components with the two-dimensional
recession cones, one unbounded component between the tentacles
with the one-dimensional recession cone and the bounded component.
The other two amoebas depicted in Fig.~\ref{img:amoeba12}~(right)
and~\ref{img:amoeba34}~(left) exhibit five connected components in
their respective complements and topologically assume an
intermediate position between the solid and the optimal amoebas
defined by polynomials with the Newton polygon~$\mathcal{N}$.
\end{example}

The existing analytic methods~\cite{FPT,Rullgaard} do not in
general allow one to predict the topological type of~a~polynomial
with generic coefficients. From the computational point of~view
the tasks of~depicting the amoebas in Fig.~\ref{img:amoeba12}
and~\ref{img:amoeba34} are rather similar. Yet, to predict the
existence of~a~bounded connected component of~a~given
order~\cite{FPT} in an amoeba complement by means of~analytic
methods is in general a~task of~formidable
complexity~\cite{Purbhoo}.

\section{COMPUTING TWO-DIMENSIONAL AMOEBAS}

\begin{definition}\rm We will call the \textit{<<carcase>> of~an amoeba}~$\mathcal{A}$
any subset of~$\mathcal{A}$, such that the number of~connected
components of~the complement to the intersection~$\mathcal{A}\cap
B$ for a~sufficiently large ball~$B$ is as big as it could
possibly be (that is, equal to the number of~connected components
in the complement of~$\mathcal{A}$ in~$\mathbb{R}^n$).
\end{definition}

We remark that the carcase of~an amoeba is not uniquely defined.
However, the topology of~its complement in a~sufficiently big ball
is well-defined and is as complex as possible. When speaking
of~depicting an amoeba we will mean depicting its suitable
carcase.

In this section we consider bivariate polynomials of~the form
$p\left(x,y\right)=\sum c_{ij}x^{i}y^{j}$, $i=0,\ldots,m$,
$j=0,\ldots,n$ and provide an algorithm for computing their
amoebas.

Theorem~\ref{3thmfptestimate} yields that the geometry of~the
amoeba~$\mathcal{A}_{p}$ is closely related to the properties
of~the Newton polytope~$\mathcal{N}_{p}$ of~the polynomial~$p$.
Yet, the coefficients of~$p$ also play a~role and determine the
size of~the carcase of~the amoeba in question. In what follows,
the boundary of~the domain where the carcase of~an amoeba is
depicted has been determined experimentally.

The next straightforward algorithm which computes the amoeba of a
bivariate polynomial has been used by numerous authors in various
forms~\cite{Forsberg,Nilsson,Johansson}. We include it for the
sake of completeness and future reference.

\begin{algorithm} \caption{Algorithm for computing the amoeba~$\mathcal{A}_{p}$
of a~bivariate polynomial}\label{alg:Amoeba2}

\begin{algorithmic}[1]

\Require List of~the polynomial coefficients  \textit{cx\_list},
in $x$ by all monomials $y^{k}$, $k=0,\ldots,\deg_{y}p$; the
boundaries of~the rectangular domain in the logarithmic amoeba
space $a$, $b$; the number of~values of~the absolute value $n_{r}$
and the angle~$n{}_{\varphi}$ of~the complex variable.

\Ensure List of~coordinates of~points that belong to the amoeba
carcase $z\_list$, $w\_list$.

\Procedure{Amoeba2D}{\textit{cx\_list}, $a$, $b$, $n_{r}$,
$n_{\varphi}$}

\State \textit{z\_list} := empty list

\State \textit{w\_list} := empty list

\State \textit{d} := the number of~elements in \textit{cx\_list}
-- 1

\State 1\textit{d} := $\left(1,\ldots,1\right)\in\mathbb{R}^{d}$
\Comment{the vector with $d$ units}

\State $h_{r}:=\left(\exp\left(b\right)-\exp\left(a\right)\right)/\left(n_{r}-1\right)$

\State $h_{\varphi}:=2\pi/\left(n_{\varphi}-1\right)$

\For{$r=\exp(a): \exp(b): h_r$}

\For{$\varphi=0: 2\pi: h_\varphi$}

\State$x:=r*\exp\left(\sqrt{-1}*\varphi\right)$

\State$y:=\mathrm{roots}(cx\_list)$ \Comment{the vector of~zeros
of the polynomial with the coefficients $cx\_list$}

\State Add $\mathrm{Log}(\left|x\right|*1d)$ to $z\_list$

\State Add $\mathrm{Log}(\left|y\right|)$ to $w\_list$

\EndFor

\EndFor

\State\textbf{return} $\left\{ z\_list,w\_list\right\} $

\EndProcedure

\end{algorithmic}

\end{algorithm}

To obtain a~picture of~good quality the steps of~the
Algorithm~\ref{alg:Amoeba2} are repeated with the variables~$x$
and~$y$ interchanged. The points with the computed coordinates are
depicted in the same figure.

\begin{example}\rm For the polynomial~$p_{4}\left(x,y\right)$
in Example~\ref{exm:imgAmoeba} the lists of~polynomial
coefficients in the variables $x$ and $y$ are as follows:
\textit{cx\_list}=$\{1,\:3+4x+x^{2},\:1+x\}$ and
\textit{cy\_list}=$\{y,\:1+4y,\:1+3y+y^{2}\}$. The (carcase
of~the) amoeba is depicted in the rectangle
$\left[-5,5\right]\times\left[-5,5\right]$. The number of~values
of the absolute value of~a~variable is $n_{r}=2000$ while the
number of~values of~its argument is $n{}_{\varphi}=180$.
Fig.~\ref{img:amoeba12} and Fig.~\ref{img:amoeba34} feature the
amoebas of~the polynomials in Example~\ref{exm:imgAmoeba} computed
by means of~Algorithm~\ref{alg:Amoeba2}.

\begin{figure}[ht!]
\begin{centering}
\includegraphics[width=0.9\columnwidth]{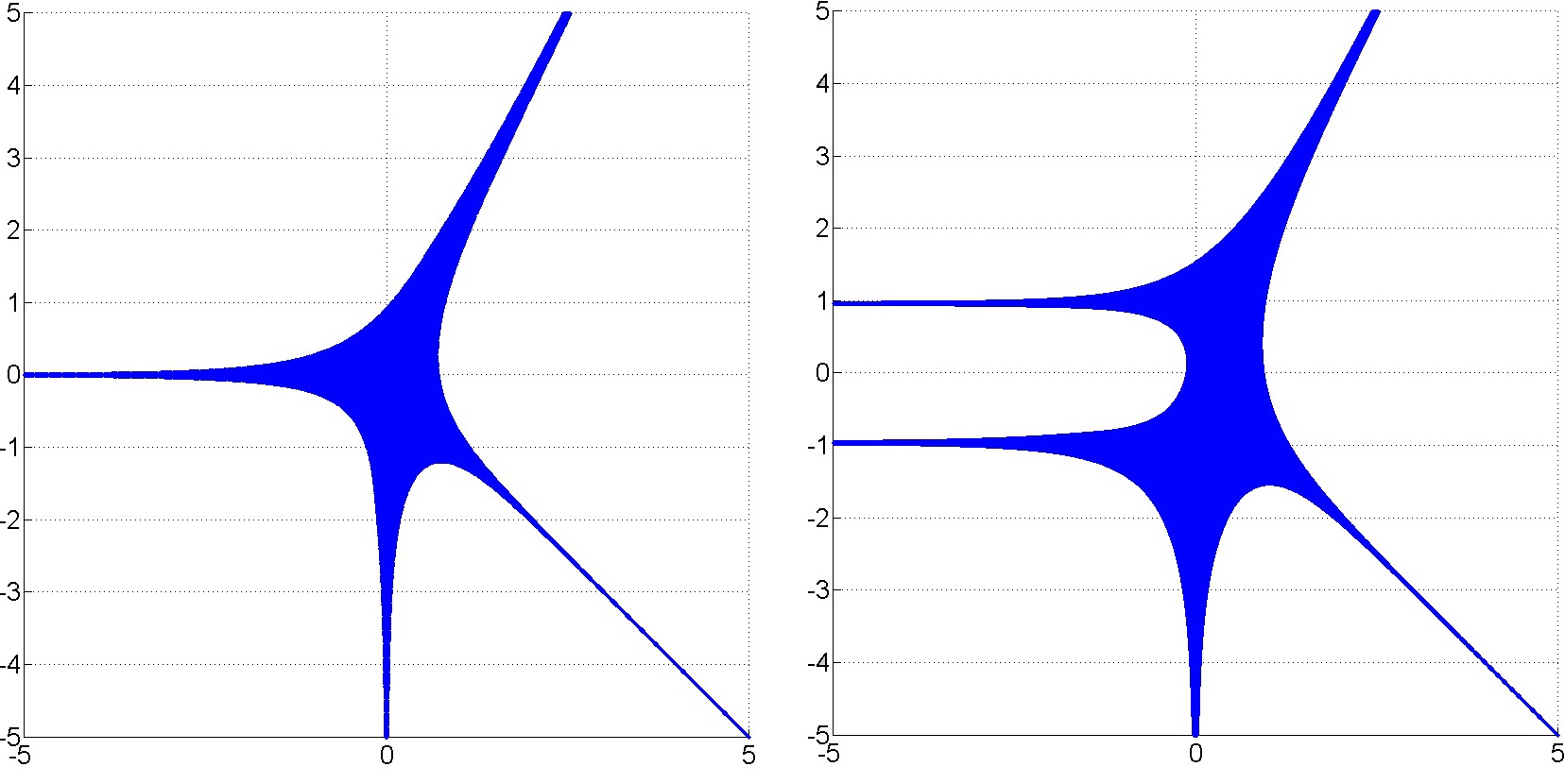}
\par\end{centering}

\caption{The amoebas of~the polynomials $p_{1}\left(x,y\right)$
and $p_{2}\left(x,y\right)$} \label{img:amoeba12}
\end{figure}

\begin{figure}[ht!]
\begin{centering}
\includegraphics[width=0.9\columnwidth]{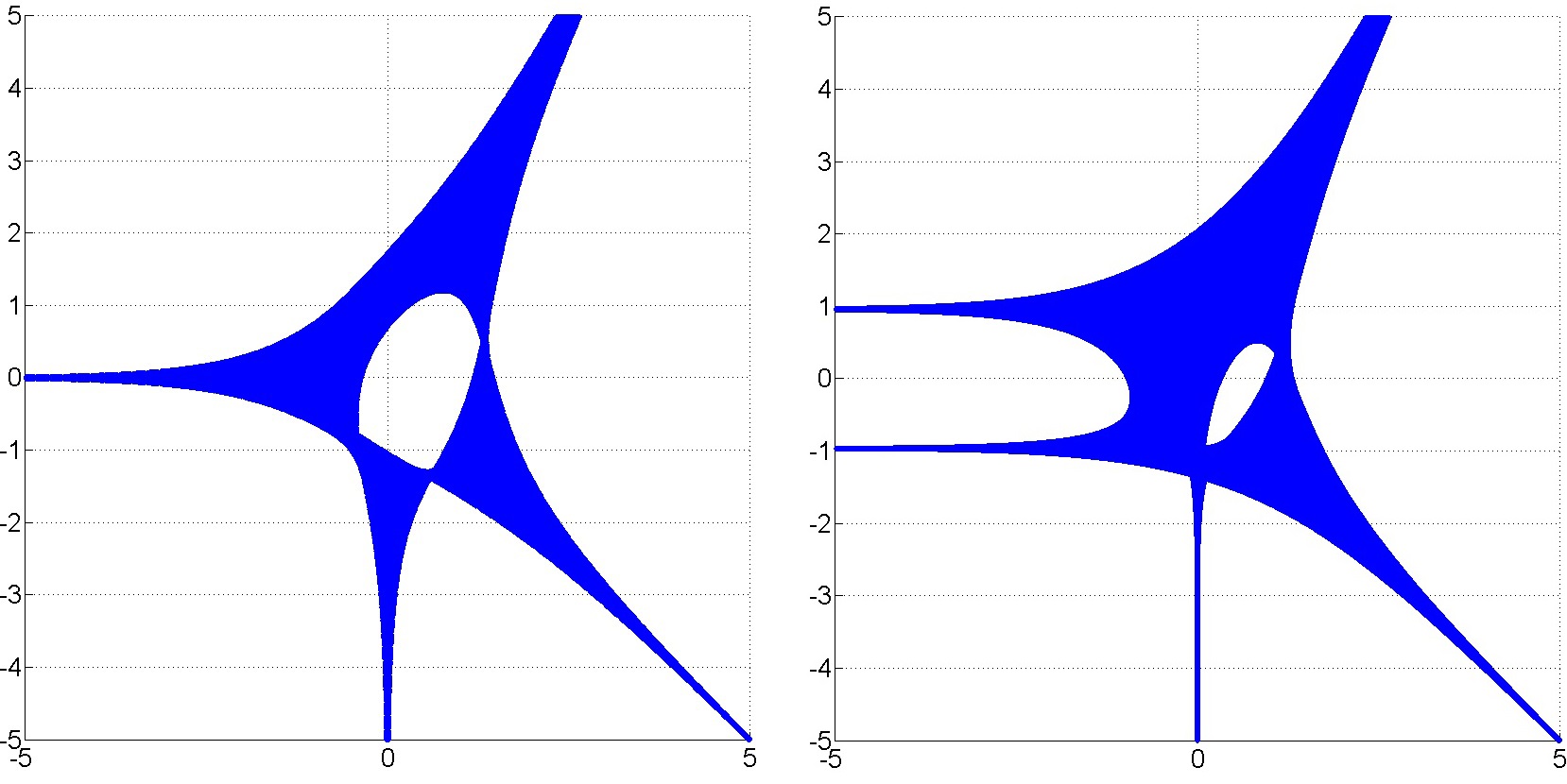}
\par\end{centering}

\caption{The amoebas of~the polynomials $p_{3}\left(x,y\right)$
and $p_{4}\left(x,y\right)$} \label{img:amoeba34}
\end{figure}

\end{example}

The topologically more involved amoeba of~the polynomial in
Example~\ref{exm:big_optimal_amoeba} has been computed in
a~similar way.

\section{GENERATING OPTIMAL POLYNOMIALS}\label{optimal_poly}

In this section we employ the notion of~a~hypergeometric
polynomial for the purpose of~constructive generation of~optimal
amoebas. We will need the following auxiliary definition.

\begin{definition}\rm\label{def:HGfunction}
A formal Laurent series
\begin{equation}
\sum_{s\in\mathbb{Z}^{n}}\varphi\left(s\right)\,x^{s}\label{series}
\end{equation}
is called \textit{hypergeometric} if for any $j=1,\ldots,n$ the
quotient $\varphi(s+e_{j})/\varphi(s)$ is a~rational function
in~$s = (s_1,\ldots,s_n).$ Throughout the paper we denote this
rational function by $P_{j}(s)/Q_{j}(s+e_{j}).$
Here~${\{e_j\}}_{j=1}^{n}$ is the standard basis of~the
lattice~$\mathbb{Z}^n.$ By the \textit{support} of~this series we
mean the subset of~$\mathbb{Z}^n$ on which $\varphi(s)\neq 0.$
\end{definition}

A \textit{hypergeometric function} is a~(multi-valued) analytic
function obtained by means of~analytic continuation
of~a~hypergeometric series with a~nonempty domain of~convergence
along all possible paths.

\begin{theorem} {\rm (Ore, Sato, cf~\cite{Sadykov})}\label{thm:Ore-Sato}
The coefficients of~a~hypergeometric series are given by the
formula
\begin{equation}
\varphi\left(s\right)=t^{s}\,U\left(s\right)\,
\prod_{i=1}^{m}\Gamma\left(\left\langle {\bf A}_{i},s\right\rangle
+c_{i}\right),\label{oresatocoeff}
\end{equation}
where $t^{s}=t_{1}^{s_{1}}\ldots t_{n}^{s_{n}}$,
$t_{i},c_{i}\in\mathbb{C}$, ${\bf
A}_{i}=\left(A_{i,1},\ldots,A_{i,n}\right)\in\mathbb{Z}^{n}$,
$i=1,\ldots,m$ and $U\left(s\right)$ is the product of~a~certain
rational function and a~periodic function $\phi\left(s\right)$
such that $\phi\left(s+e_{j}\right)=\phi\left(s\right)$ for every
$j=1,\ldots,n$. \end{theorem}

Given the above data ($t_{i},c_{i},{\bf A}_{i},U\left(s\right)$)
that determines the coefficient of~a~multivariate hypergeometric
Laurent series, it is straightforward to compute the rational
functions $P_{i}\left(s\right)/Q_{i}\left(s+e_{i}\right)$ using
the $\Gamma$-function identity.

\begin{definition}\rm\label{def:Horn system}
A (formal) Laurent series
$\sum_{s\in\mathbb{Z}^{n}}\varphi\left(s\right)x^{s}$ whose
coefficient satisfies the relations
$\varphi(s+e_{j})/\varphi\left(s\right)=
P_{j}\left(s\right)/Q_{j}\left(s+e_{j}\right)$ is a~(formal)
solution to the following system of~partial differential equations
of hypergeometric type:
\begin{equation}
x_{j}P_{j}\left(\theta\right)f\left(x\right)=
Q_{j}\left(\theta\right)f\left(x\right),\,\,\,j=1,\ldots,n.\label{eq:horn}
\end{equation}
Here $\theta=\left(\theta_{1},\ldots,\theta_{n}\right)$,
$\theta_{j}=x_{j}\frac{\partial}{\partial x_{j}}$.

The system~(\ref{eq:horn}) will be referred to as \textit{the Horn
hypergeometric system defined by the Ore-Sato
coefficient~$\varphi(s)$} (see~\cite{DS}) and denoted by ${\rm
Horn}\left(\varphi\right)$.\end{definition}

In the present paper we will only consider the important special
case of~an Ore-Sato coefficient~(\ref{oresatocoeff}) with
$t_{i}=1$ for all $i=1,\ldots,n$ and $U\left(s\right)\equiv1$. The
Horn system associated with such an Ore-Sato coefficient will be
denoted by ${\rm Horn}\left(A,c\right)$, where~$A$ is the matrix
with the rows ${\bf A}_{1},\ldots,{\bf A}_{m}\in\mathbb{Z}^{n}$
and $c=\left(c_{1},\ldots,c_{m}\right)\in\mathbb{C}^{m}$. In this
case the operators $P_{j}\left(\theta\right)$ and
$Q_{j}\left(\theta\right)$ explicitly determine the
system~(\ref{eq:horn}):
\[
P_{j}\left(s\right)=\prod_{i:A_{i,j}>0}
\prod_{\ell_{j}^{(i)}=0}^{A_{i,j}-1}\left(\left\langle {\bf
A}_{i},s\right\rangle +c_{i}+\ell_{j}^{\left(i\right)}\right),
\]
\[
Q_{j}\left(s\right)=\prod_{i:A_{i,j}<0}
\prod_{\ell_{j}^{(i)}=0}^{|A_{i,j}|-1}\left(\left\langle {\bf
A}_{i},s\right\rangle +c_{i}+\ell_{j}^{\left(i\right)}\right).
\]

For a~convex integer polytope~$\mathcal{N},$ we construct the
list~$B$ of~the outer normals to the faces of~$\mathcal{N}.$ We
denote the length of~this list by~$q$. We assume the elements
of~$B$ to be normalized so that the coordinates of~each outer
normal are integer and relatively prime. Define the Ore-Sato
coefficient
\begin{equation}
\varphi\left(s\right)= \frac{1}{\prod\limits
_{j=1}^{q}\Gamma\left(1-\left\langle B_{j},s\right\rangle
-c_{j}\right)}.\label{eq:calcOreSato}
\end{equation}
The hypergeometric system~(\ref{eq:horn}) defined by the Ore-Sato
coefficient $\varphi(s)$ admits a~polynomial solution with several
interesting properties. Under some additional assumptions it turns
out to have an optimal amoeba. We will employ the following
algorithm to generate optimal polynomials of~hypergeometric type.

\begin{algorithm}
\caption{Generation of~a~bivariate hypergeometric polynomial and
its defining system of~equations}\label{alg:Hyperpoly2D}

\begin{algorithmic}[1]

\Require List of~vertices~$N\_list$ of~a~convex integer
polygon~$P$.

\Ensure List of~coefficients $c\_list$ of~the hypergeometric
polynomial $p\left(x,y\right)$ whose Newton polygon is~$P$.

\Procedure{Hyperpoly2D}{\textit{N\_list}}

\State$B$ = list of~outer normals to the sides of~$P$.

\State$\varphi\left(s,t\right)=1/\prod\limits
_{j=1}^{q}\Gamma\left(1-\left\langle
B_{j},\left(s,t\right)\right\rangle -c_{j}\right)$

\State$c\_list$ = list of~the coefficients of~the hypergeometric
polynomial

\State$R_{1}=\mathrm{FunctionExpand}[\varphi\left(s+1,t\right)/\varphi\left(s,t\right)]$

\State$R_{2}=\mathrm{FunctionExpand}[\varphi\left(s,t+1\right)/\varphi\left(s,t\right)]$

\State$P_{1}=\mathrm{Numerator}[R_{1}]$

\State$P_{2}=\mathrm{Numerator}[R_{2}]$

\State$Q_{1}=\mathrm{Denumerator}[R_{1}]$

\State$Q_{2}=\mathrm{Denumerator}[R_{2}]$

\State$\theta_{x}=x\frac{dp}{dx}$

\State$\theta_{y}=y\frac{dp}{dy}$

\State $p$ = the polynomial defined by $c\_list$

\If{$x P_1 (\theta)p=Q_1(\theta)p$ \textbf{and} $y
P_2(\theta)p=Q_2(\theta)p$}

\State\textbf{return} $\left\{ c\_list\right\} $

\EndIf

\EndProcedure

\end{algorithmic}

\end{algorithm}

\begin{example}\rm\label{ex:HGExample}
The outer normals (normalized as explained above) to the sides
of~the polygon shown in Fig.~\ref{img:Newton} are as follows:
$\left(0,-1\right)$, $\left(-1,0\right)$, $\left(1,2\right)$,
$\left(1,-1\right)$. Using~(\ref{eq:calcOreSato}) we obtain
\[
\begin{array}{c}
\varphi\left(s,t\right)=
(\Gamma\left(s+1\right)\Gamma\left(t+1\right)\Gamma\left(-s-2t+5\right)\Gamma\left(-s+t+2\right))^{-1}.
\end{array}
\]
By Definition~\ref{def:Horn system} the above Ore-Sato coefficient
$\varphi\left(s,t\right)$ gives rise to the polynomials
\[
\begin{array}{c}
P_{1}\left(s,t\right)=\left(-s-2t+4\right)\left(-s+t+1\right),\quad
Q_{1}\left(s,t\right)=s+2,\\
P_{2}\left(s,t\right)=\left(s+2t-4\right)\left(s+2t-3\right),\quad
Q_{2}\left(s,t\right)=\left(t+2\right)\left(-s+t+3\right).
\end{array}
\]
The corresponding hypergeometric system is defined by the linear
partial differential operators
\begin{equation}
\left\{ \begin{array}{l}
x\,\left(-\theta_{x}-2\theta_{y}+4\right)\left(-\theta_{x}+\theta_{y}+1\right)-\left(\theta_{x}+2\right),\\
y\,\left(\theta_{x}+2\theta_{y}-4\right)\left(\theta_{x}+2\theta_{y}-3\right)-
\left(\theta_{y}+2\right)\left(-\theta_{x}+\theta_{y}+3\right).
\end{array}\right.\label{hornSystemExample}
\end{equation}
It is straightforward to check that the hypergeometric polynomial
$p(x,y)=1+4x+6y+24xy+12x^{2}y+2y^{2}$ belongs to the kernels
of~the operators~(\ref{hornSystemExample}).
\end{example}

\begin{example}\rm\label{exm:big_optimal_amoeba}
Using Algorithms~\ref{alg:Amoeba2} and~\ref{alg:Hyperpoly2D} we
compute the coefficients of~the hypergeometric polynomial
supported in the polygon~$\mathcal{N}_{p_{h}\left(x,y\right)}$
shown in Fig.~\ref{img:big_amoeba}~(left):
$p_{h}\left(x,y\right)=2x^{2}+20xy+72x^{2}y+20x^{3}y+5y^{2}+160xy^{2}+
450x^{2}y^{2}+160x^{3}y^{2}+5x^{4}y^{2}+12y^{3}+300xy^{3}+800x^{2}y^{3}+
300x^{3}y^{3}+12x^{4}y^{3}+5y^{4}+160xy^{4}+450x^{2}y^{4}+160x^{3}y^{4}+
5x^{4}y^{4}+20xy^{5}+72x^{2}y^{5}+20x^{3}y^{5}+2x^{2}y^{6}.$ The
amoeba~$\mathcal{A}_{p_{h}\left(x,y\right)}$ of~this polynomial is
depicted in Fig.~\ref{img:big_amoeba}~(right).

A more involved hypergeometric polynomial is given by
$p_{h2}(x,y)=-456456x^{3}+488864376x^{2}y-28756728x^{3}y+
25420947552x^{2}y^{2}-244432188x^{3}y^{2}+3003x^{4}y^{2}-
119841609888xy^{3}+127104737760x^{2}y^{3}-465585120x^{3}y^{3}+
6006x^{4}y^{3}+1396755360y^{4}-508418951040xy^{4}+139815211536x^{2}y^{4}-
232792560x^{3}y^{4}+1729x^{4}y^{4}+4190266080y^{5}-355893265728xy^{5}+
41611670100x^{2}y^{5}-29628144x^{3}y^{5}+57x^{4}y^{5}+698377680y^{6}-
58663725120xy^{6}+3328933608x^{2}y^{6}-705432x^{3}y^{6}-2327925600xy^{7}+
55023696x^{2}y^{7}-16930368xy^{8}.$ Its defining Ore-Sato
coefficient equals
$\varphi(s,t)=\Gamma(s+t-4)\Gamma(-4s+t-16)\Gamma(-3s-2t-5)\Gamma(3s-t-3)\Gamma(2s+t-5).$
The amoeba of~this polynomial is shown in
Fig.~\ref{img:biggest_amoeba}.

\begin{figure}[ht!]
\begin{centering}
\includegraphics[width=0.9\columnwidth]{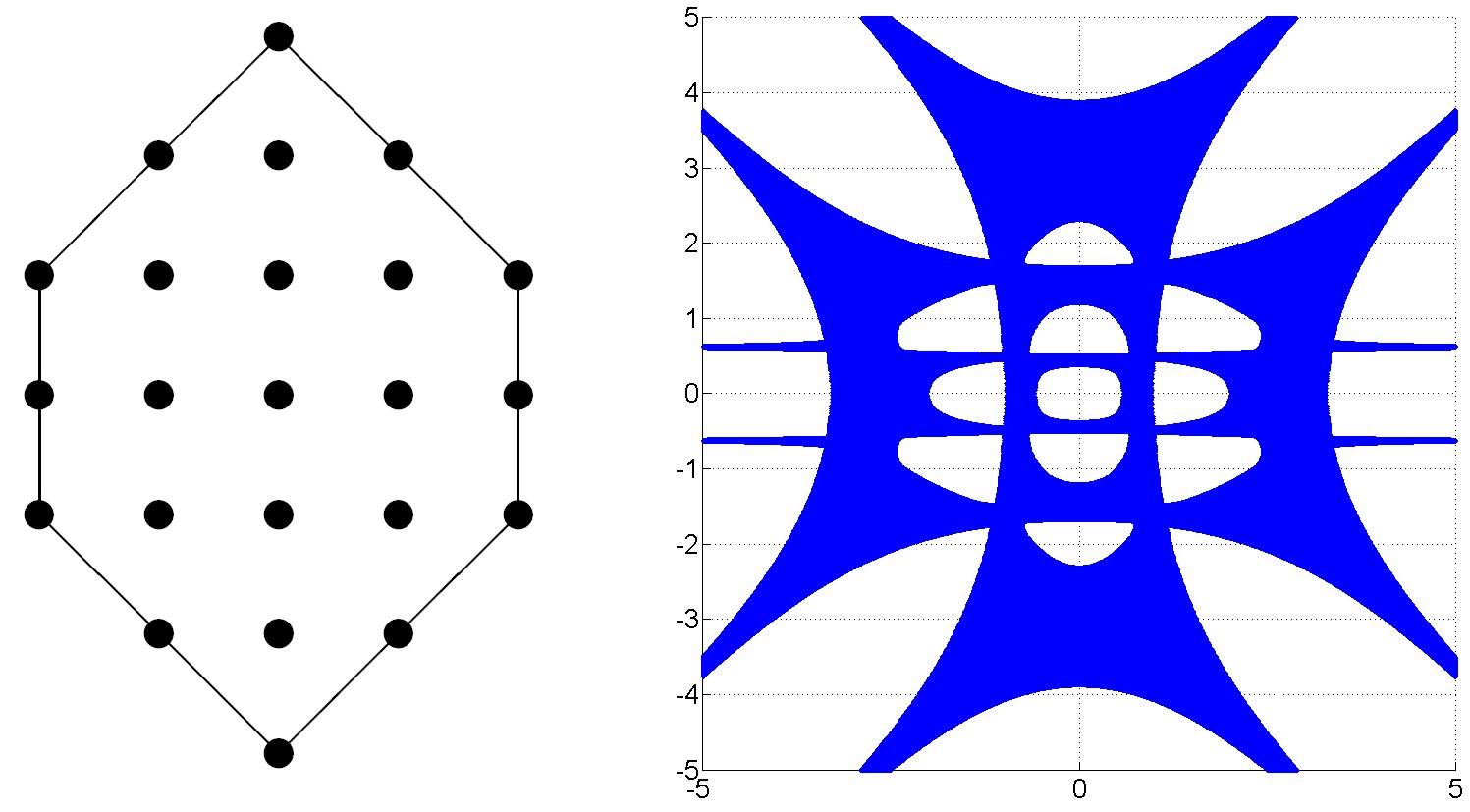}
\par\end{centering}

\caption{The Newton polytope and the amoeba of~the hypergeometric
polynomial $p_{h}\left(x,y\right)$} \label{img:big_amoeba}
\end{figure}

\begin{figure}[ht!]
\begin{centering}
\includegraphics[width=0.9\columnwidth]{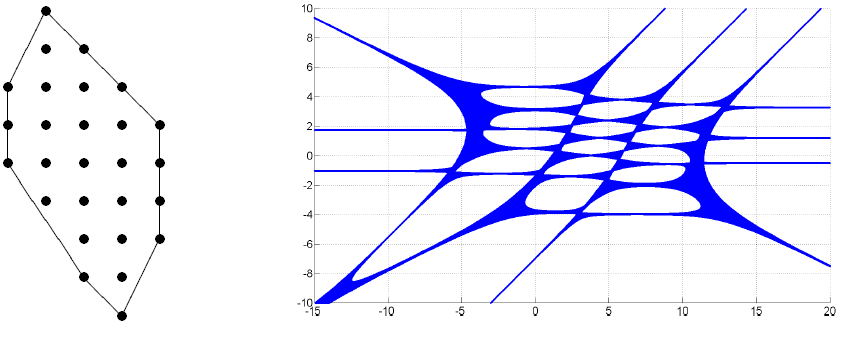}
\par\end{centering}

\caption{The Newton polytope and the amoeba of~the hypergeometric
polynomial $p_{h2}\left(x,y\right)$} \label{img:biggest_amoeba}
\end{figure}

\end{example}

\section{COMPUTING CONTOURS OF~AMOEBAS OF~BIVARIATE POLYNOMIALS}

The boundary of~an amoeba admits a~rich analytic structure that is
revealed in the following definition.

\begin{definition}\rm The \textit{contour} of~the amoeba~$\mathcal{A}_{p\left(x\right)}$
is the set~$\mathcal{C}_{p\left(x\right)}$ of~critical points
of~the logarithmic map~$\mathrm{Log}$ restricted to the zero locus
of the polynomial~$p\left(x\right)$:
\[
\mathrm{Log}:\;\left\{ p\left(x\right)=0\right\} \rightarrow\mathbb{R}^{n}.
\]
\end{definition}

The structure of~the contour of~an amoeba can be described in
terms of~the logarithmic Gauss map which is defined as follows.

\begin{definition}
\rm Denote by~$\mathrm{Gr}(n,k)$ the Grassmannian
of~$k$-dimensional subspaces in~$\mathbb{C}^{n}$. \textit{The
logarithmic Gauss map}~\cite{Kapranov} is defined to be the map
$\gamma:\;\mathcal{H}\longmapsto\mathrm{Gr}(n,k)$ that maps
a~regular point $x\in\mathrm{reg}\thinspace\mathcal{H}$ into the
normal subspace~$\gamma(x)$ to the
image~$\mathrm{log}\thinspace\mathcal{H}$.
\end{definition}

The boundary of~an amoeba $\partial\mathcal{A}_{p\left(x\right)}$
is necessarily a~subset of~the
contour~$\mathcal{C}_{p\left(x\right)}$ but is in general
different from it.

Knowing the structure of~the contour of~an amoeba is important for
describing the topological structure of~the amoeba complement.
Experiments show that a~cusp of~the contour inside the <<body>>
of~the corresponding amoeba is a~counterpart of~a~missing
connected component in its complement.

One of~the ways to draw the contour of~an amoeba is to depict the
solutions to the system of~algebraic equations
\begin{equation}
\begin{cases}
p\left(x,y\right)=0,\\
x\frac{\partial p\left(x,y\right)}{\partial x}-uy\frac{\partial p\left(x,y\right)}{\partial y}=0.
\end{cases}\label{eq:contour}
\end{equation}

Here $u\in\mathbb{R}\cup\left\{ \infty\right\}$ is a~real
parameter that encodes the slope of~the normal line to the contour
of the amoeba.

Eliminating the variables~$x$ and~$y$ out of~the
system~(\ref{eq:contour}) we obtain polynomials
$s\left(y,u\right)$ and $t\left(x,u\right)$ that can be used to
depict the contour of~the amoeba by means of~the following
algorithm.

\begin{algorithm} \caption{Computing the contour~$\mathcal{C}_{p}$
of the amoeba of~a~bivariate polynomial}\label{alg:Countor2D}

\begin{algorithmic}[1]

\Require List of~polynomial coefficients \textit{cx\_list}
\textit{cy\_list} depending on $u$, by all monomials $x^{k}$,
$k=0,\ldots,\deg_{x}t$, $y^{m}$, $m=0,\ldots,\deg_{y}s$; initial
value $u_{1}$, final value $u_{n}$ and the step~$h_{u}$.

\Ensure List of~coordinates of~points $z\_list$, $w\_list$ that
belong to the contour of~the amoeba.

\Procedure{Contour2D}{\textit{cx\_list}, \textit{cy\_list}, $u_1$, $u_n$, $h_u$}

\State \textit{z\_list} := the empty list

\State \textit{w\_list} := the empty list

\State \textit{dx }:= the number of~elements in \textit{cx\_list}
-- 1

\State \textit{dy }:= the number of~elements in \textit{cy\_list}
-- 1

\For{$u=u_1: u_n: h_u$}

\State$x:=\mathrm{roots}(cx\_list)$ \Comment{the vector of~zeros
of the polynomial with the coefficients $cx\_list$}

\State$y:=\mathrm{roots}(cy\_list)$ \Comment{the vector of~zeros
of the polynomial with the coefficients $cy\_list$}

\State Add $\mathrm{Log}(\left|x\right|)$ to $z\_list$

\State Add $\mathrm{Log}(\left|y\right|)$ to $w\_list$

\EndFor

\State\textbf{return} $\left\{ z\_list,w\_list\right\} $

\EndProcedure

\end{algorithmic}

\end{algorithm}

\begin{remark}\rm
The lists of~coordinates that are obtained at each iteration
of~the cycle in the above algorithm might contain additional
elements that do not correspond to the amoeba contour. They are
sorted out by checking against the system of~equations
(\ref{eq:contour}).\end{remark}

\begin{example}\rm
The system of~equations~(\ref{eq:contour}) associated with the
first polynomial in Example~\ref{exm:imgAmoeba} has the form
\[
\begin{cases}
1+x+y+xy+y^{2}+x^{2}y=0,\\
x-uy+xy-uxy+2x^{2}y-ux^{2}y-2uy^{2}=0.
\end{cases}
\]

Eliminating variables~$x$ and~$y$ yields the equations
$u-2uy+u^{2}y-4uy^{2}+2u^{2}y^{2}-2uy^{3}-u^{2}y^{3}+3uy^{4}+
4uy^{5}+u^{2}y^{5}=1-y-4y^{2}-9y^{3}-7y^{4}-4y^{5}$ and
$-3u^{2}-3ux-5u^{2}x-2ux^{2}+u^{2}x^{2}+3ux^{3}+5u^{2}x^{3}+2ux^{4}+
3u^{2}x^{4}+ux^{5}+u^{2}x^{5}=x^{2}+2x^{3}+5x^{4}+2x^{5}$
respectively. The lists of~the coefficients that depend on~$u$ are
\textit{cx\_list}~=~$\{-1+u,\:1-2u+u^{2},\:4-4u+2u^{2},\:9-2u-u^{2},\:7+3u,4+4u+u^{2}\}$
and
\textit{cy\_list}~=~$\{-3u^{2},\:-3u-5u^{2},\:-1-2u+u^{2},\:-2+3u+5u^{2},\:-5+2u+3u^{2},\:-2+u+u^{2}\}$.
The parameter~$u$ assumes values in the interval
$\left[-120,120\right]$ with the step $h_{u}=0,001$. The obtained
contours of~the amoebas of~the polynomials in
Example~\ref{exm:imgAmoeba} have nontrivial structure and are
shown in Fig.~\ref{img:countor12} and~\ref{img:countor34}.
\begin{figure}[ht!]
\begin{centering}
\includegraphics[width=0.9\columnwidth]{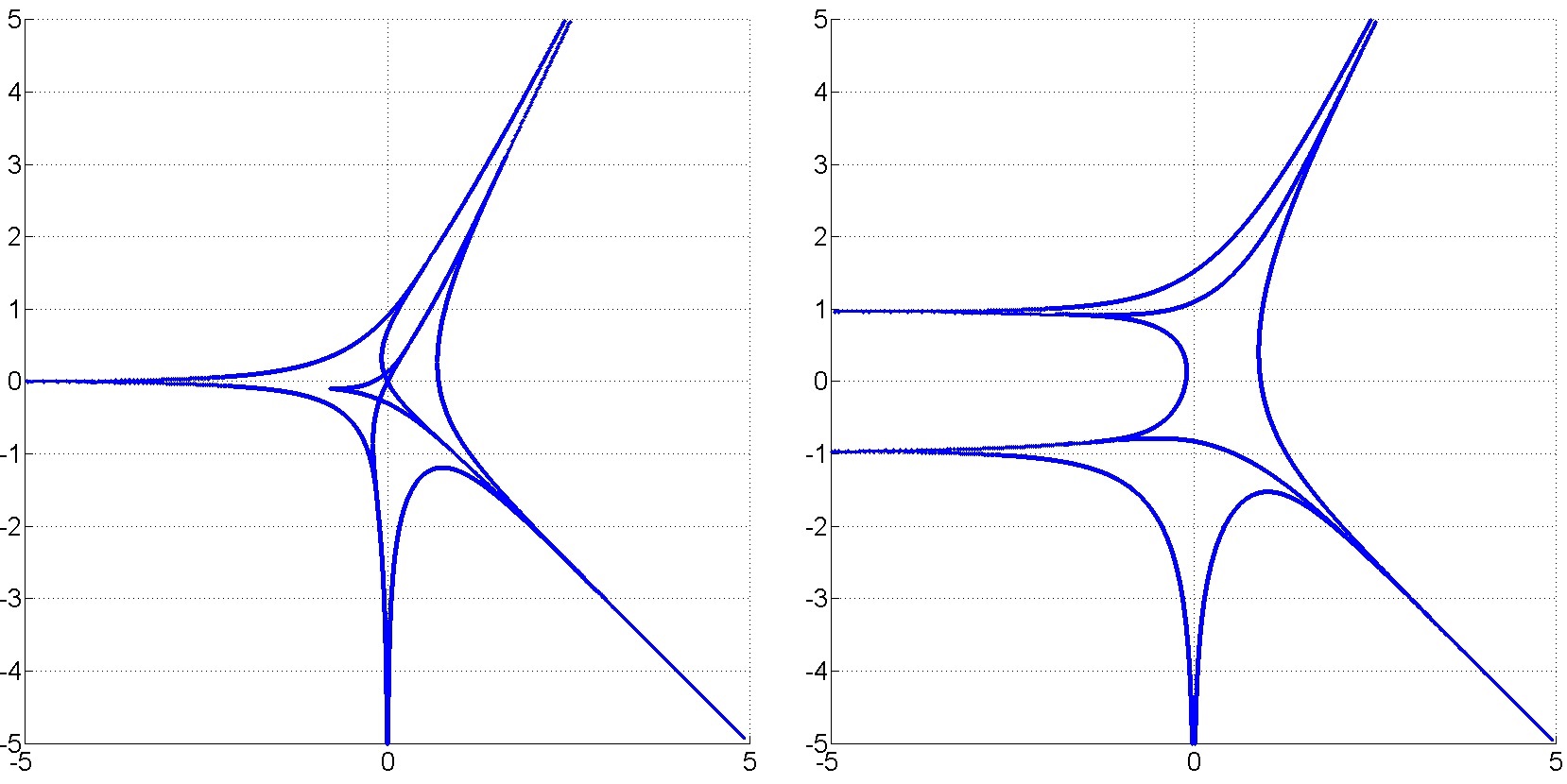}
\par\end{centering}

\caption{Contours of~the amoebas of~the polynomials
$p_{1}\left(x,y\right)$ and $p_{2}\left(x,y\right)$}
\label{img:countor12}
\end{figure}

\begin{figure}[ht!]
\begin{centering}
\includegraphics[width=0.9\columnwidth]{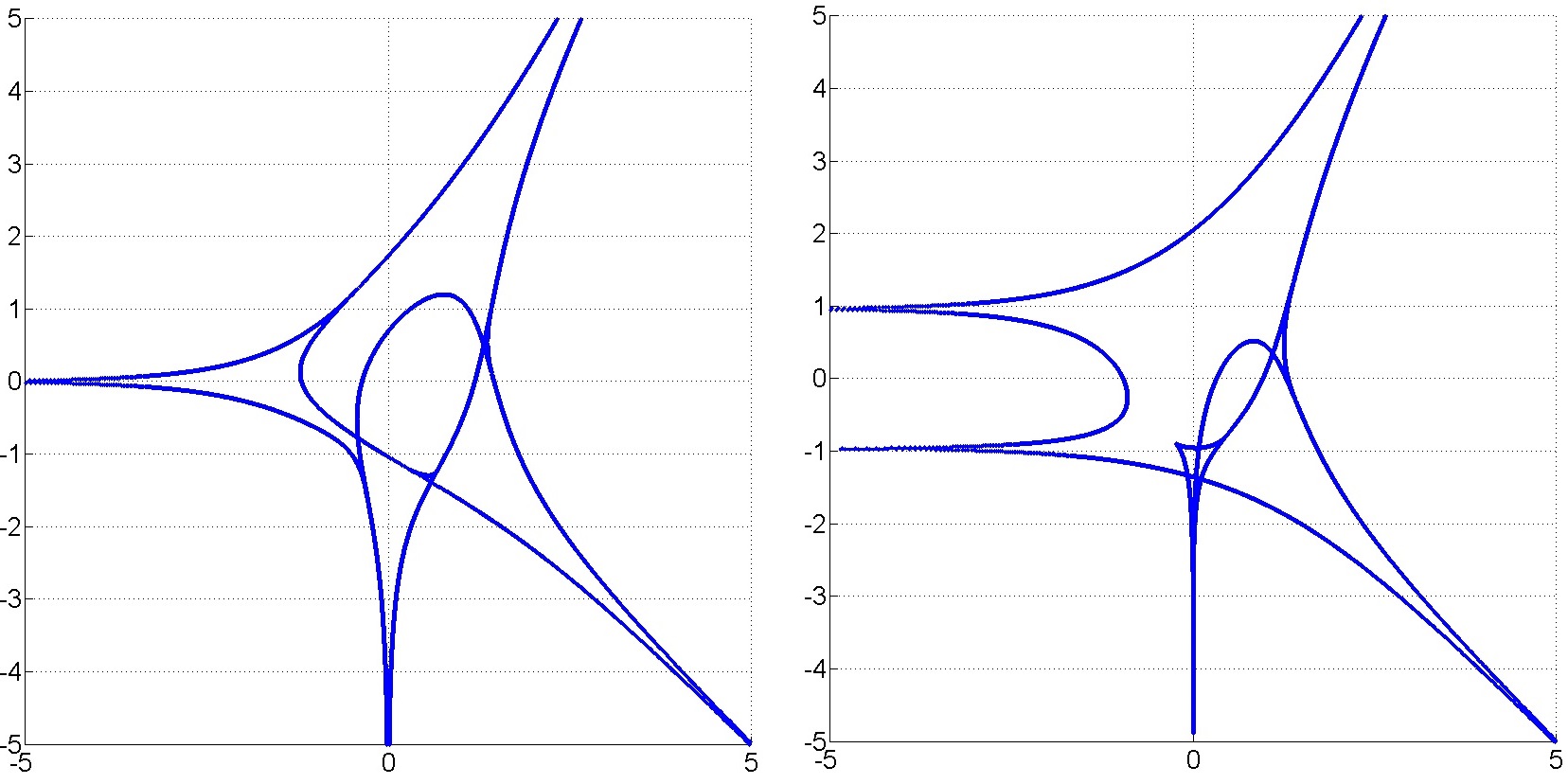}
\par\end{centering}

\caption{Contours of~the amoebas of~the polynomials
$p_{3}\left(x,y\right)$ and $p_{4}\left(x,y\right)$}
\label{img:countor34}
\end{figure}
\end{example}

\section{COMPUTING TWO-DIMENSIONAL COMPACTIFIED AMOEBAS}

\begin{definition}\rm\textit{The compactified amoeba}~$\overline{\mathcal{A}}_{p\left(x\right)}$
of a~polynomial~$p\left(x\right)$ is defined to be the image
of~its zero locus under the mapping
\begin{equation}
M\left(x\right)=\frac{\sum\limits_{\alpha\in
A}\alpha\cdot\left|x^{\alpha}\right|}{\sum\limits _{\alpha\in
A}\left|x^{\alpha}\right|} 
=\frac{\sum\limits_{\left(\alpha_{1},\ldots,\alpha_{n}\right)\in
A}\left(\alpha_{1},\ldots,\alpha_{n}\right)\cdot\left|x^{\alpha_{1}}\cdot\ldots\cdot
x^{\alpha_{n}}\right|}{\sum\limits
_{\left(\alpha_{1},\ldots,\alpha_{n}\right)\in
A}\left|x^{\alpha_{1}}\cdot\ldots\cdot
x^{\alpha_{n}}\right|}.\label{eq:moment_map}
\end{equation}
\end{definition}
Numeric computation of~two-dimensional compactified amoebas is
similar to that of~their affine counterparts
(see~Algorithm~\ref{alg:Amoeba2}). The main computational issue is
dealing with the moment map~(\ref{eq:moment_map}) instead of~the
logarithmic map into the affine space.

\begin{example}\rm
Applying~(\ref{eq:moment_map}) to the polynomials in
Example~\ref{exm:imgAmoeba} yields the mapping
\[
\left(x,y\right)\longmapsto\mu\left(x,y\right)=
\]
\[
\frac{(1\cdot\left(0,0\right)+\left|x\right|\cdot\left(1,0\right)+\left|y\right|\cdot\left(0,1\right)
+\left|xy\right|\cdot\left(1,1\right)+\left|y^{2}\right|\cdot\left(0,2\right)+\left|x^{2}y\right|\cdot\left(2,1\right))}
{\left(\left|1\right|+\left|x\right|+\left|y\right|+\left|xy\right|+\left|y^{2}\right|+\left|x^{2}y\right|\right)}=
\]
\[
\frac{(\left|x\right|+\left|xy\right|+2\left|x^{2}y\right|,\left|y\right|+\left|xy\right|+2\left|y^{2}\right|+\left|x^{2}y\right|)}
{\left(\left|1\right|+\left|x\right|+\left|y\right|+\left|xy\right|+\left|y^{2}\right|+\left|x^{2}y\right|\right)}.
\]

The corresponding compactified amoebas inside the Newton polygon
of their defining polynomials are shown in
Fig.~\ref{img:compact12} and~\ref{img:compact34}.

\begin{figure}[ht!]
\begin{centering}
\includegraphics[width=0.9\columnwidth]{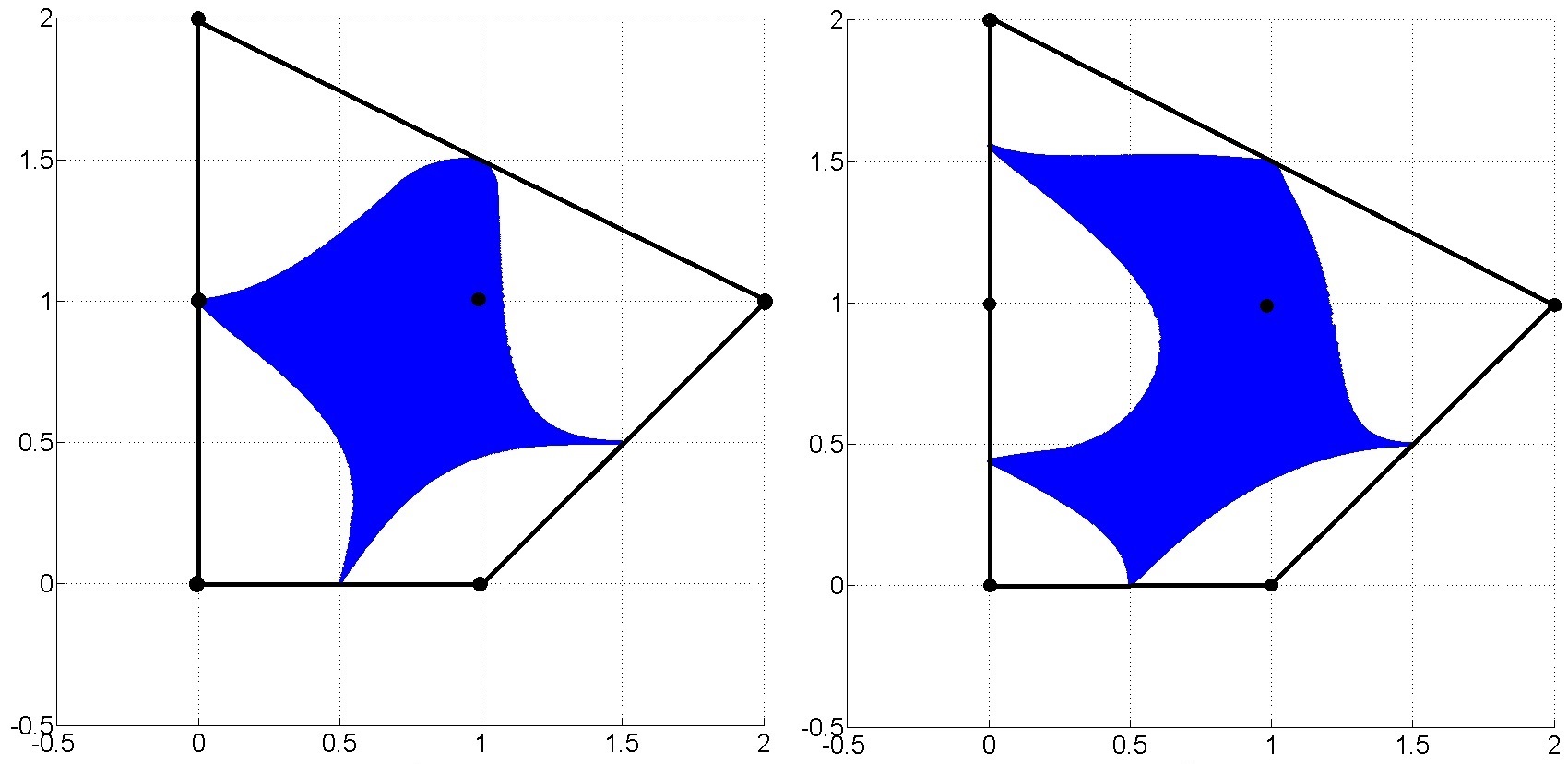}
\par\end{centering}

\caption{Compactified amoebas of~the polynomials
$p_{1}\left(x,y\right)$ and $p_{2}\left(x,y\right)$ in their
Newton polygon} \label{img:compact12}
\end{figure}

\begin{figure}[ht!]
\begin{centering}
\includegraphics[width=0.9\columnwidth]{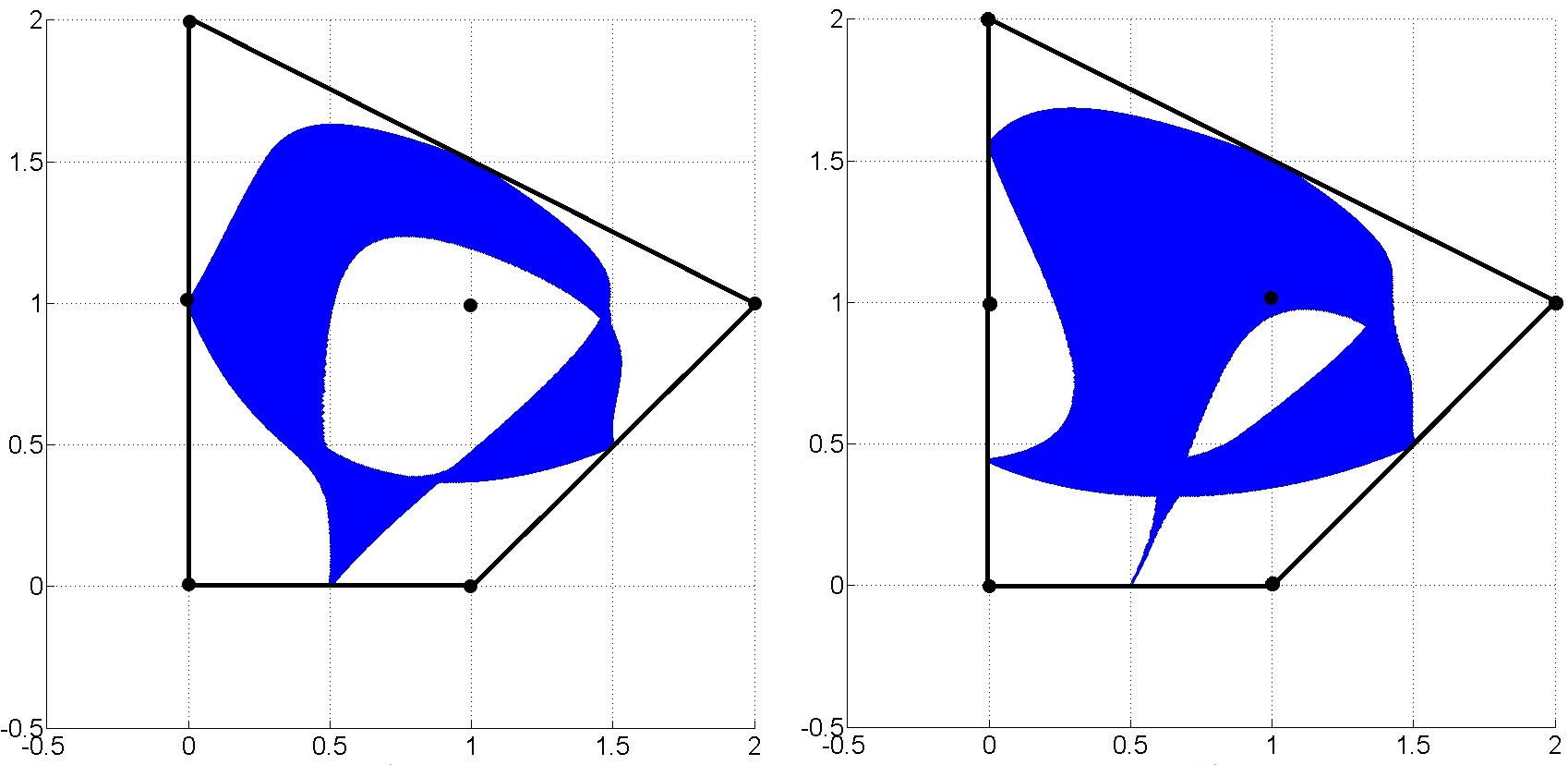}
\par\end{centering}

\caption{Compactified amoebas of~the polynomials
$p_{3}\left(x,y\right)$ and $p_{4}\left(x,y\right)$ in their
Newton polygon} \label{img:compact34}
\end{figure}
\end{example}

\section{MULTIVARIATE OUTLOOK}

Depicting three-dimensional amoebas represents a~substantial
computational challenge due to the complex geometry of~their
shape. We will now treat the problem of~computing sections
of~amoebas of~polynomials in three variables by two-dimensional
hyperplanes.

Consider a~polynomial in three complex variables
$p\left(x,y,z\right)=\sum c_{ijk}x^{i}y^{j}z^{k}$, $i=0,\ldots,m$,
$j=0,\ldots,n$, $k=0,\ldots,s$. To compute the section of~its
amoeba by the plane $\left|z\right|=\mathrm{const}$ we fix the
absolute value of~$z$ and modify Algorithm~\ref{alg:Amoeba2} by
adding a~cycle with respect to the values of~the argument
$\varphi_{z}\in\left[0;2\pi\right]$.

\begin{algorithm}
\caption{Computing section of~the amoeba $\mathcal{A}_{p}$
of~a~polynomial in three variables}\label{alg:Amoeba3D}

\begin{algorithmic}[1]

\Require List of~polynomial coefficients \textit{cxz\_list},
depending on $x$ and $z$ by all monomials $y^{k}$,
$k=0,\ldots,\deg_{y}p$; bounds $a$, $b$ for the rectangular domain
in the amoeba space where the amoeba section is depicted; the
numbers~$n_{r}$ and~$n{}_{\varphi}$ of~values of~the absolute
value and the argument of~the variable $z\in\mathbb{C},$
respectively.

\Ensure List of~points that belong to the amoeba section
$z\_list$, $w\_list$.

\Procedure{Amoeba3D}{\textit{cxz\_list}, $a$, $b$, $n_{r}$, $n_{\varphi}$, $z$}

\State \textit{z\_list} := empty list

\State \textit{w\_list} := empty list

\State \textit{d} := the number of~elements in \textit{cxz\_list}
-- 1

\State 1\textit{d} := $\left(1,\ldots,1\right)\in\mathbb{R}^{d}$
\Comment{the vector of~$d$ units}

\State $h_{r}:=\left(\exp\left(b\right)-\exp\left(a\right)\right)/\left(n_{r}-1\right)$

\State $h_{\varphi}:=2\pi/\left(n_{\varphi}-1\right)$

\For{$r=\exp(a): \exp(b): h_r$}

\For{$\varphi=0: 2\pi: h_\varphi$}

\State$x:=r*\exp\left(\sqrt{-1}*\varphi\right)$

\State$y:=\mathrm{roots}(cx\_list)$ \Comment{the vector of~zeros
of the polynomial with the coefficients $cx\_list$}

\State Add $\mathrm{Log}(\left|x\right|*1d)$ to $z\_list$

\State Add $\mathrm{Log}(\left|y\right|)$ to $w\_list$

\EndFor

\EndFor

\State\textbf{return} $\left\{ z\_list,w\_list\right\} $

\EndProcedure

\end{algorithmic}

\end{algorithm}

\begin{example}\rm Using the above Algorithm~\ref{alg:Amoeba3D}
we compute the section of~the amoeba of~the polynomial
$p\left(x,y,z\right)=1+3y+y^{2}+6xy+x^{2}y+xyz+xyz^{2}$. This
polynomial is one of~the simplest polynomials with tetrahedral
Newton polytopes that contain integer points in the interior as
well as in the relative interior of~faces of~all positive
dimensions (see Fig.~\ref{img:Newton_amoeba3}).
\begin{figure}[ht!]
\begin{center}
\begin{picture}(170,180)
\put(50,80){\circle*{5}} \put(100,80){\circle*{5}}
\put(150,80){\circle*{5}} \put(80,50){\circle*{5}}
\put(60,20){\circle*{5}} \put(80,100){\circle*{5}}
\put(80,150){\circle*{5}} \put(50,80){\line(1,0){10}}
\put(50,80){\line(2,5){30}} \put(150,80){\line(-1,1){70}}
\put(150,80){\line(-3,-2){90}} \put(50,80){\line(1,-6){10}}
\put(80,150){\line(-1,-6){22}}
\put(50,80){\vector(0,1){100}} \put(75,80){\vector(1,0){95}}
\put(50,80){\vector(-1,-2){40}} \put(165,90){$x$}
\put(55,175){$z$} \put(0,0){$y$}
\end{picture}
\end{center}
\caption{The Newton polytope $\mathcal{N}_{p\left(x,y,z\right)}$}
\label{img:Newton_amoeba3}
\end{figure}
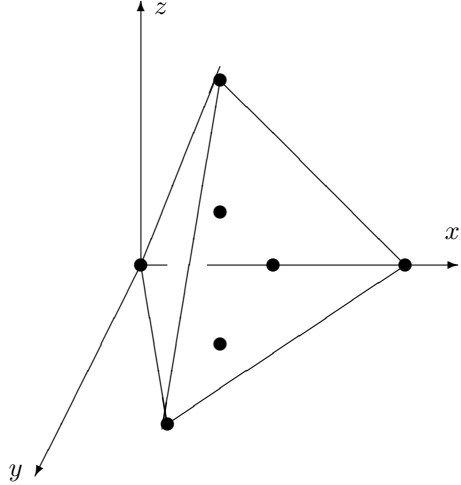

\begin{figure}[ht!]
\begin{centering}
\includegraphics[width=0.5\columnwidth]{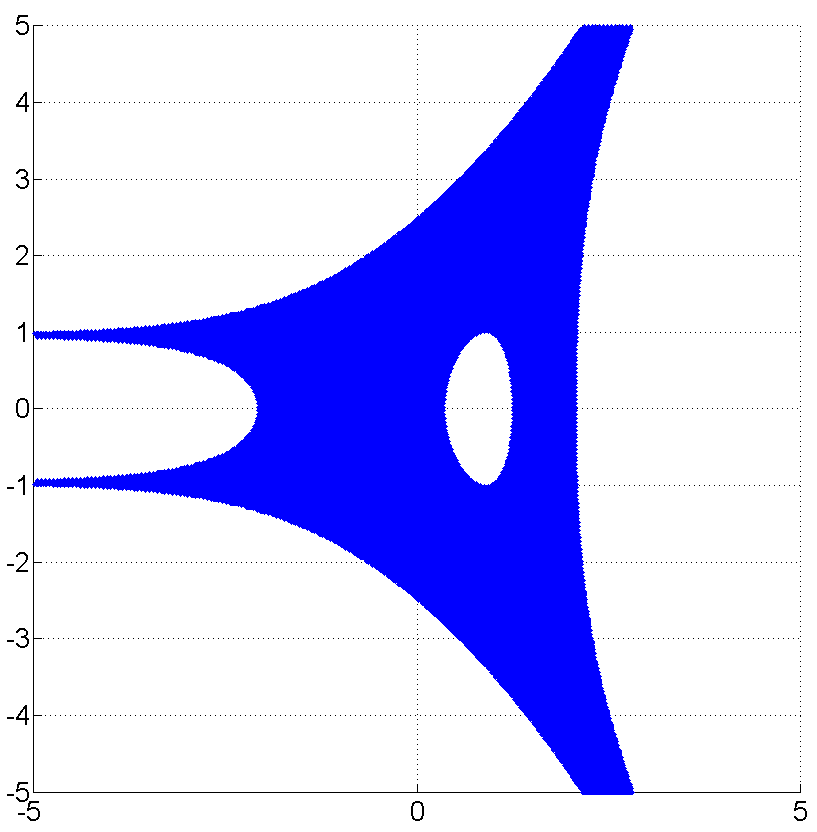}
\par\end{centering}

\caption{The section of~the amoeba of~the polynomial
$p\left(x,y,z\right)$ by the plane $\log|z|=5$}
\label{img:amoeba3}
\end{figure}
The section of~the amoeba $\mathcal{A}_{p\left(x,y,z\right)}$ by
the plane $\log\left|z\right|=5$ is shown in
Fig.~\ref{img:amoeba3}.

\end{example}

\begin{example}\rm
We now consider a~computationally more challenging optimal
hypergeometric polynomial in three variables supported in
a~regular integer octahedron. Due to symmetry it suffices to
consider its part that belongs to the positive orthant and is
shown in Fig.~\ref{img:3d_amoeba_optimal}~(left). We use
a~three-dimensional version of~Algorithm~\ref{alg:Hyperpoly2D} to
compute the corresponding (uniquely determined up to scaling)
hypergeometric polynomial:
$p_{h}\left(x,y,z\right)=x^{2}y^{2}+36x^{2}yz+36xy^{2}z+256x^{2}y^{2}z+
36x^{3}y^{2}z+36x^{2}y^{3}z+x^{2}z^{2}+36xyz^{2}+256x^{2}yz^{2}+36x^{3}yz^{2}+
y^{2}z^{2}+256xy^{2}z^{2}+1296x^{2}y^{2}z^{2}+256x^{3}y^{2}z^{2}+x^{4}y^{2}z^{2}+
36xy^{3}z^{2}+256x^{2}y^{3}z^{2}+36x^{3}y^{3}z^{2}+x^{2}y^{4}z^{2}+36x^{2}yz^{3}+
36xy^{2}z^{3}+256x^{2}y^{2}z^{3}+36x^{3}y^{2}z^{3}+36x^{2}y^{3}z^{3}+x^{2}y^{2}z^{4}$.
Fig.~\ref{img:3d_amoeba_optimal}~(right) shows the intersection
of~the amoeba~$\mathcal{A}_{p_{h}\left(x,y,z\right)}$ with the
three coordinate hyperplanes as well as with the hyperplanes
$|x|=3,$ $|x|=8$, $|y|=6$, $|z|=4$ in the logarithmic amoeba
space. The figure space is orthogonal to the vector~$(1,1,1)$.

\begin{figure}[ht!]
\begin{centering}
\includegraphics[width=0.9\columnwidth]{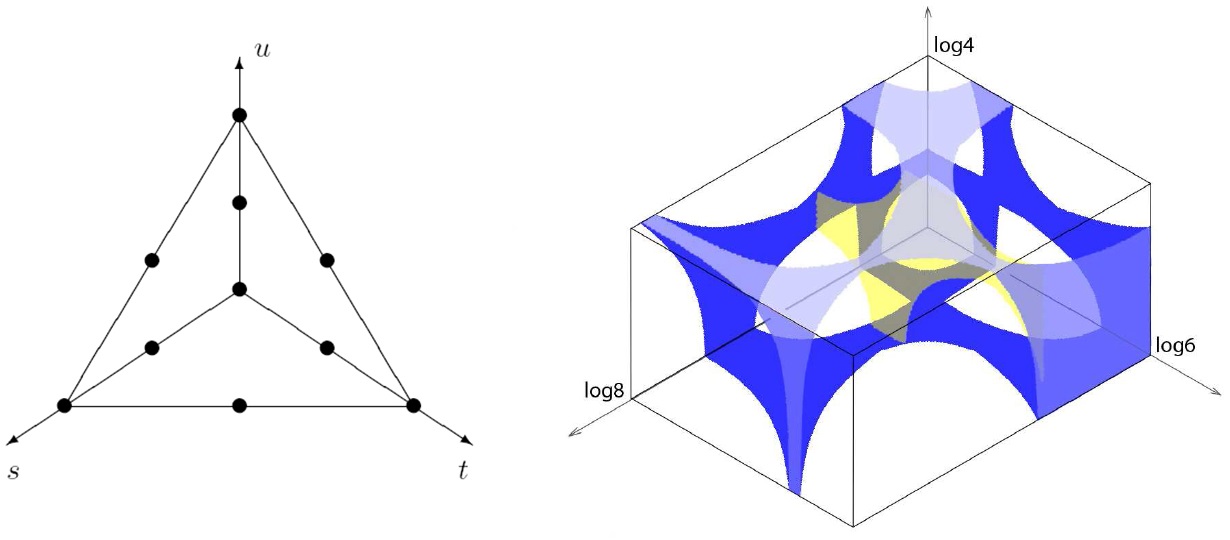}
\par\end{centering}

\caption{The part of~the support of~$p_{h}\left(x,y,z\right)$ that
belongs to the positive orthant and the carcase of~the
corresponding part of~the amoeba} \label{img:3d_amoeba_optimal}
\end{figure}

\end{example}

\noindent \textbf{Acknowledgements.} The research was partially
supported by the state order of~the Ministry of~Education and
Science of~the Russian Federation for Siberian Federal University
(task 1.1462.2014/K, the second author), by Government of~the
Russian Federation (project no. 14.Y26.31.0006, the third author).
and by the Russian Foundation for Basic Research, projects
15-31-20008-mol\_a\_ved (the second and the third authors),
15-01-00277-a and 14-01-00283-a (the second author).

\end{document}